\documentclass[titlepage]{article}
\usepackage{graphicx}        
\usepackage{hyperref}  

\begin{document}

\title{Big Data and Fog Computing \thanks{To Appear as a contribution in \emph{Encyclopedia of Big Data Technologies}, Sherif Sakr and Albert Zomaya eds., Springer Nature, 2018}}
\author{Yogesh Simmhan\\
Indian Institute of Science, Bangalore, India\\\href{mailto:simmhan@iisc.ac.in}{simmhan@iisc.ac.in}}

\maketitle

\section{Synonyms} 

Cloudlets

\section{Definitions} 

\begin{description}
\item[Fog Computing]{A model of distributed computing comprising of virtualized, heterogeneous, commodity computing and storage resources for hosting applications, analytics, content and services, that are accessible with low latency from the edge of wide area networks where clients are present, while also having back-end connectivity to cloud computing resources.}
\end{description}

\section{Background}
Ever since computers could connect over a network, computing paradigms have under-gone cyclical phases on \emph{where} within the network the computation is performed. While the original mainframes till the 1970's were large, centralized time-sharing machines accessed by multiple users through remote terminals, the Personal Computers (PCs) of the 1980's heralded local processing for individuals~\cite{bell-1986}. The growth of Local-Area Networks (LAN), the Internet, and the World Wide Web (WWW) brought about client-server models in the 1990's where many clients could access content hosted on individual servers, though most of the processing was still done on the PC~\cite{sinha-1992}. Web services and eCommerce of the 2000's lead to the growth of \emph{cloud computing}, where computation once again skewed to centralized data centers, but with server-farms rather than single servers hosting services that were consumed by PCs~\cite{hwang-2013}. A complementary phenomenon in that decade was Peer-to-Peer (P2P) systems where PCs distributed across the Internet would work collaboratively for content sharing, to tackle bandwidth limitations of the Internet~\cite{milojicic-2002}. Both cloud computing and P2P signaled the arrival of the \emph{Big Data} age, where the ability to collect large enterprise, scientific and web datasets and media content put an emphasis on being able to share and process them at large scales.

The decade of 2010 is seeing a similar cyclical shift, but at a faster pace due to several technology advances. Starting with a more centralized cloud computing model hosting thousands of Virtual Machines (VMs), we have seen the roll-out of pervasive broadband Internet and cellular network communication combined with the rapid growth of smart phones as general-purpose computing platforms backed by cloud computing. \emph{Internet of Things (IoT)} is yet another paradigm, enabled by the convergence of these other technologies~\cite{gubbi-2013}. Sensors and constrained devices connected to the Internet are being deployed to support vertical IoT domains such as personal fitness using wearables, smart utilities using metering infrastructure, and even self-driving cars. Both smart phones and IoT mark the advent of \emph{Edge Computing} (or mobile cloud computing). Here, ubiquitous devices numbering in the billions are present at the edge of the Wide-Area Network (WAN) that is the Internet, and host applications that either operate locally or serve as light-weight clients that publish data to or consume content from cloud services~\cite{lopez-2015,fernando-2013}.

\begin{figure*}[!t]
  \centering
	\includegraphics[width=0.95\textwidth]{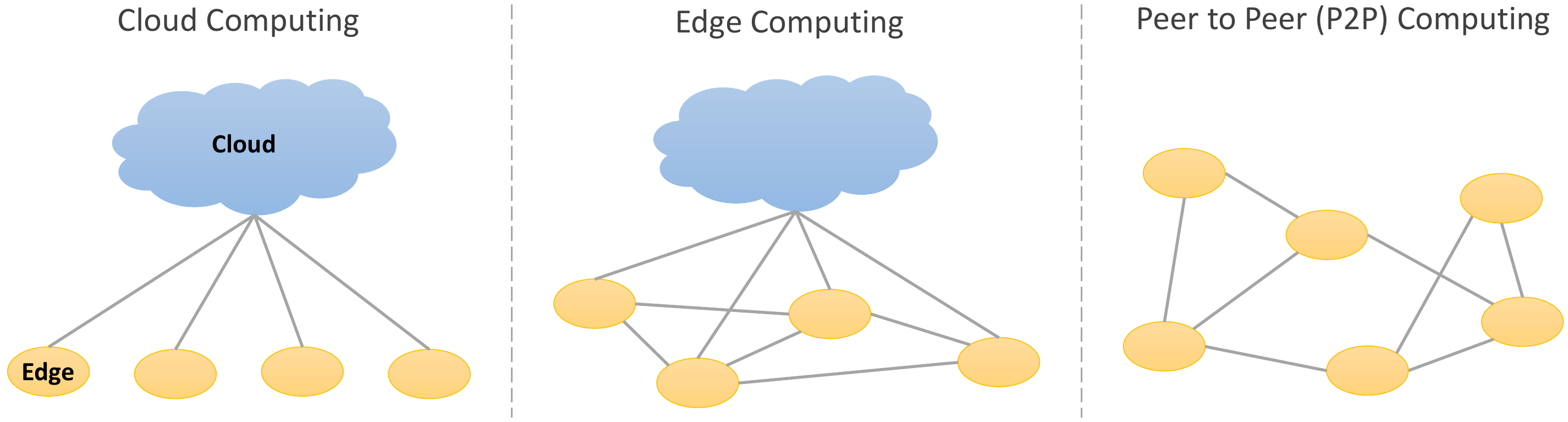}
	\caption{Connectivity between resources in Cloud, Edge and P2P models.}
    \label{fig:cloud-edge}
\end{figure*}

The limitations of an edge-only or a cloud-only model was recognized by Satyanarayanan, et al, who introduced the concept of \emph{Cloudlets}~\cite{satyanarayanan-2009}. These are resource-rich servers, relative to edge devices, that could host VMs, while also being closer in the network topology to the edge devices, relative to cloud data centers. They are designed to overcome the constrained resources available on edge platforms while reducing the network latency expended in being tethered to the cloud for interactive applications. \emph{Fog computing} generalizes (and popularizes) the notion of cloudlets.

The term ``Fog Computing'' was coined by Cisco, and first appears publicly in a talk by Flavio Bonomi, Vice President and Fellow at Cisco Systems, as part of the \emph{Network-Optimized Computing at the Edge Of the Network Workshop}, co-located with International Symposium on Computer Architecture (ISCA), in 2011~\cite{bonomi-2011}. This was further described as extending the concept of cloud computing to the edge of the network to support low-latency and geo-distributed applications for mobile and IoT domains~\cite{bonomi-2014}. Since then, fog computing has been evolving as a concept and covers a wide class of resources that sit between the edge devices and cloud data centers on the network topology, have capacities that fall between edge devices and commodity clusters on clouds, and may be managed \emph{ad hoc} as a smart gateway or professionally as a computing infrastructure~\cite{varshney-2017,vaquero-2014,yi-2015}.

\section{Motivations}

Fog computing is relevant in the context of \emph{wide-area distributed systems}, with many clients at the edge of the Internet~\cite{donnet-2007}. Such clients may be mobile or consumer devices (\emph{e.g.,} smart phone, smart watch, Virtual Reality (VR) headset) used interactively by humans, or devices that are part of IoT (\emph{e.g.,} smart meter, traffic cameras and signaling, driverless cars) for machine-to-machine (M2M) interactions. The client may serve both as \emph{consumers} of data or actuators that receive control signals (\emph{e.g.,} fitness notification on smart watch, signal change operation on traffic light), as well as \emph{producers} of data (\emph{e.g.,} heart-rate observations from smart watch, video streams from traffic cameras)~\cite{dastjerdi-2016}.

There are two contemporary models of computing for applications and analytics that use data from, or generate content and controls for, such clients at the edge of the Internet~\cite{simmhan-2017}, as illustrated in Fig.~\ref{fig:cloud-edge}. In a \emph{cloud-centric model}, also called cloud computing, data from these clients is sent to a central data-center where the processing is done and the responses, if any, are sent back to the same or different client(s)~\cite{simmhan:cise:2013,cai-2016}. In an \emph{edge-centric model}, or edge-computing, part (or even all) of the processing is done at the data source with the rest done on the cloud~\cite{beck-2014,anand-2017}. 

These have their relative advantages. Cloud computing outsources computing infrastructure management to providers, who offer elastic access to seemingly infinite compute resources on-demand which can be rented by the minute. They are also cheaper due to economies of scale at centralized locations~\cite{cai-2016}. Edge computing leverages the compute capacity of existing captive devices and reduces network transfers, both of which lower the costs. There may also be enhanced trust and context available closer to the edge~\cite{lopez-2015}.

While fog computing is still maturing, there are many reason why its rise is inevitable due to the gaps in these two common computing approaches. The \emph{network latency} from the edge client to the cloud data-center is high and variable, averaging between $20-250~ms$ depending on the location of the client and data center~\cite{he-2013}. The \emph{network bandwidth} between the edge and the cloud, similarly, averages at about $800-1200~KB/s$. Both these mean that latency sensitive or bandwidth intensive applications will offer poor performance using a cloud-centric model due to the round-trip time between edge and cloud~\cite{satyanarayanan-2009,satyanarayanan-2015}. Another factor is the connectivity of devices to the Internet. Mobile devices may be out of network coverage periodically, and cause cloud-centric applications to degrade or loose functionality~\cite{shi-2016}.

Edge computing, while avoiding issues of network time, suffer from operating on constrained devices that have limited battery, compute and memory capacities~\cite{barbera-2013}. This reduces application performance and limits sustained processing that can drain the battery. These devices may also be less robust and their network connectivity less available~\cite{aral-2017}.

Fog computing serves as a computing layer that sits between the edge devices and the cloud in the network topology. They have more compute capacity than the edge but much less so than cloud data centers. They typically have high uptime and always-on Internet connectivity. Applications that make use of the fog can avoid the network performance limitation of cloud computing while being less resource constrained than edge computing. As a result, they offer a useful balance of the current paradigms.

\section{Characteristics}
\begin{figure*}[!t]
  \centering
	\includegraphics[width=0.95\textwidth]{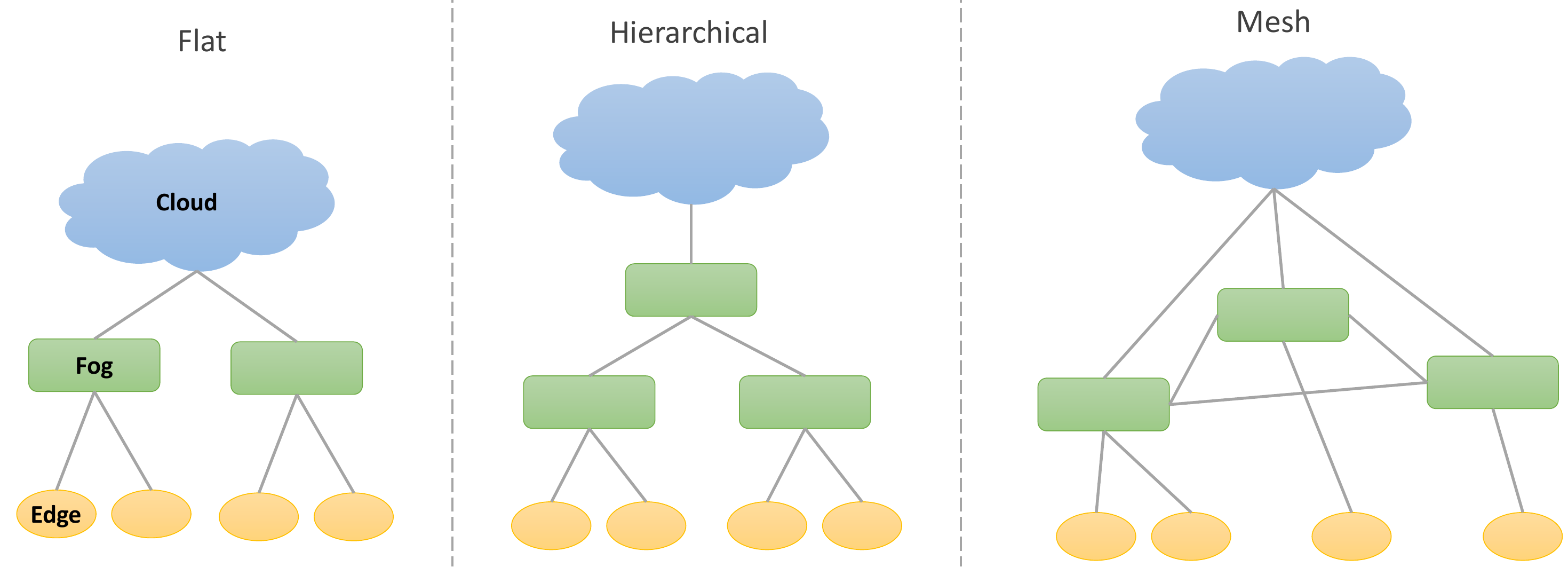}
	\caption{Different interaction models in fog computing.}
    \label{fig:fog}
\end{figure*}

Fog computing is an evolving model of computing. Hence, its perceived characteristics are broad, based on diverse view-points of the research community and industry~\cite{varshney-2017}. It is generally agreed that a defining characteristic of fog computing is its smaller network distance from the edge, but multiple \textbf{network topology architectures} exist, as shown in Fig.~\ref{fig:fog}~\cite{aazam-2014,he-2017,hou-2016}. Some models consider there to be a \emph{flat} fog layer that sits, say, at 1-hop from the edge devices and allows connectivity to the cloud. Each fog device may be responsible for a collection of edge devices that connect to it. This zone of responsibility may be based on logical grouping or spatial regions. Others consider a \emph{hierarchical} approach where fog devices form layers, where each layer is further away from the edge device and responsible for fog/edge devices that fall within its sub-tree, with the cloud forming the root. Alternatively, there are \emph{mesh} designs where fog resources can communicate with each other as well as the cloud, with each edge device assigned to a single fog. A variation of this does not distinguish between edge and fog devices, and any of these may communicate with each other, or the cloud. This approaches an edge-centric or P2P model.

The \textbf{resource capacity} in the fog can vary. At one end, Raspberry Pi devices with $1~GHz$ processors, $1~GB$ RAM and $100~Mbps$ Ethernet may serve as a gateway fog resource for lower-end edge devices. On the other hand, fog resources could be provisioned as ``micro'' or ``nano'' data-centers with clusters of servers or even accelerators present~\cite{lopez-2015}. Individual fog devices can also have heterogeneous capacities~\cite{chiang-2016}. 

Fog resources tend to be more \textbf{reliable and available} than edge resources, though lacking the robust fail-over mechanisms in the cloud that is possible due to a large resource pool. This makes them well-suited to serve as a layer for persisting data and services for the short and medium term. Fog resources themselves may be deployed in a stationary environment (\emph{e.g.,} coffee shop, airport, cell tower) or in a mobile platform (\emph{e.g.,} train, cab)~\cite{chiang-2016}. This can affect the network connectivity of the fog with the cloud, in case it uses cellular networks for Internet access, and even its energy footprint~\cite{jalali-2017}.

Fog \textbf{deployment models} are still emerging. These resources may be deployed within a public or a private network, depending on its end-use. Smart city deployments may make them available to utility services within the city network~\cite{yannuzzi-2017} while retail shops and transit services can make them available to their customers. A wider deployment for public use on-demand, say, by cellular providers, cities or even cloud providers, will make it comparable to cloud computing in terms of accessibility. These have implications on the operational costs as well~\cite{vaquero-2014}.

The fog resources may be made available \textbf{as-a-service}, similar to cloud resources. These may be virtualized or non-virtualized infrastructure~\cite{bittencourt-2015}, with containers offering a useful alternative to hypervisor-based VMs that may be too heavy-weight for lower-end fog resources~\cite{anand-2017}. However, there is a still a lack of a common platform, and programming models are just emerging~\cite{hong-2013,ravindra-2017}. Most applications that use the fog tend to be custom designed, and there has only been some theoretical work on scheduling applications on edge, fog and cloud~\cite{brogi-2017,ghosh-2017}.


\section{Role of Big Data}
One of the key rationales for deploying and using fog computing is Big Data generated at the edge of the network. This is accelerated by IoT deployments. Traditional web clients which just consume services and content from the WWW saw the growth of \emph{Content Distribution Networks (CDN)} to serve these with low latency. The data from IoT sensors is instead generating data at the clients that is being pushed to the cloud~\cite{cai-2016}. In this context, fog computing has been described as acting like an inverse CDN~\cite{satyanarayanan-2015}.

A large swathe of IoT data comes as \emph{observational streams}, or time-series data, from widely distributed sensors~\cite{naas-2017,shukla-2017}. These data streams vary in their rates -- once every $15~mins$ for smart utility meters, every second for heart rate monitoring by a fitness watch to $50~Hz$ by Phasor Measurement Units (PMU) in smart power grids, and the number of sensors can range in the millions for city-scale deployments. These are \textbf{high velocity} data streams that are latency sensitive and need online analytics and decision-making to provide, say, health alerts or manage the power grid behavior~\cite{simmhan:cise:2013}. Here, fog computing can help move the decision-making close to the edge to reduce latency.

Another class of \textbf{high volume} data that is emerging is from video streams from traffic and surveillance cameras, for public safety, intelligent traffic management and even driverless cars~\cite{aazam-2016}. Here, the bandwidth consumed in moving the data from the edge to the cloud can be enormous as high-definition cameras become cheap but network capacity growth does not keep pace~\cite{satyanarayanan-2015}. The applications that need to be supported can span real-time video analytics to just recording footage for future use. Fog computing can reduce the bandwidth consumed in the core Internet and limit data movement to the local network. In addition, it can offer higher compute resources and accelerators to deploy complex analytics as well.

In addition, \textbf{telemetry data} from monitoring the health of the IoT fabric itself may form a large corpus~\cite{yannuzzi-2017}. Related to this is provenance that describes the source and processing done on the distributed devices that may be essential to determine the quality and veracity of the data. Fog can help with the collection and curation of such ancillary data streams as well.

\textbf{Data archival} is another key requirement within such applications~\cite{cai-2016,vaquero-2014}. Besides online analytics, the raw or pre-processed observational data may need to be persisted for medium or long-term to train models for Machine Learning (ML) algorithms, for auditing to justify automated decisions, or to analyze on-demand based on external factors. Depending on the duration of persistence, data may be buffered in the fog either transiently or for movement to the cloud during off-peak hours to shape bandwidth usage. Data can also be filtered or aggregated to send only the necessary subset to the cloud.

Lastly, \textbf{metadata} describing the entities in the eco-system will be essential for information integration from diverse domains~\cite{anand-2017}. These can be static or slow changing data, or even complex knowledge or semantic graphs that are constructed. They may need to be combined with real-time data to support decision making~\cite{zhou-2017}. The fog layer can play a role in replicating and maintaining this across distributed resources closer to the edge.

There has been rapid progress on Big Data Platforms on clouds and clusters, with frameworks like Spark, Storm, Flink, HBase, Pregel and TensorFlow helping store and process large data volumes, velocities, and semi-structured data. Clouds also offer these platforms as a service. However, there is a lack of programming models, platforms and middleware to support various processing patterns necessary over Big Data at the edge of the network that can effectively leverage edge, fog and cloud resources~\cite{pradhan-2017}.

\section{Examples of Applications}
The applications driving the deployment and need for fog computing are diverse. But some requirements are recurrent: low latency processing, high volume data, high computing or storage needs, privacy and security, and robustness. These span virtual and augmented reality (VR/AR) applications and gaming~\cite{yi-2015}, Industrial IoT~\cite{chiang-2016} and field support for the military~\cite{lewis-2014}. Some of the emerging and high impact applications are highlighted below.

\subsection{Smart Cities}
Smart cities are a key driver for fog computing, and these are already being deployed. The Barcelona city's ``street-side cabinets'' offer fog resources as part of the city infrastructure~\cite{yannuzzi-2017}. Here, audio and environment sensors, video cameras, and power utility monitors are packaged along-side compute resources and network back-haul capacity as part of fog cabinets placed along streets. These help aggregate data from sensors, perform basic analytics and also offer WiFi hot-spots for public use. As an example, audio analytics at the fog helps identify loud noises that then triggers a surveillance camera to capture a segment of video for further analysis. Similar efforts are underway at other cities as well~\cite{amrutur-2017}. These go toward supporting diverse smart city applications for power and water utilities, intelligent transport, public safety, etc., both by the city and by app-developers.

One of the major drivers for such city-scale fog infrastructure is likely to be \emph{video surveillance} that is starting to become pervasive in urban spaces~\cite{satyanarayanan-2015}. Such city or even crowd-sourced video feeds form \emph{meta-sensors} when combined with recent advances in deep neural networks (DNN). \emph{E.g.,} feature extraction and classification can count traffic and crowds, detect pollution levels, identify anomalies, etc. As such, they can replace myriad other environment sensors when supported by real-time analytics. Such DNN and ML algorithms are computationally cost to train and even infer, and can make use of fog resources with accelerators~\cite{khochare-2017}.


\subsection{Healthcare}
\emph{Wearables} are playing a big role in not just personal fitness but also as assistive technologies in healthcare. Projects have investigated the use of such on-person monitoring devices to detect when stroke patients have fallen and need external help~\cite{cao-2015}. Others use eye-glass cameras and head-up displays (HUDs) to offer verbal and cognitive cues to Alzheimer's patients suffering from memory loss, based on visual analytics~\cite{satyanarayanan-2009}. Predictive analytics over brain signals monitored from EEG headsets have been used for real-time mental state monitoring~\cite{sadeghi-2016}. These are then used to mitigate external conditions and stimuli that can affect patients' mental state.

All these applications require low latency and reliable analytics to be performed over observations that are being collected by wearables. Given that such devices need to be light-weight, the computation is often out-sourced to a smart phone or a server that acts as a fog resource, and to which the observational data is passed for computation. There are also decisions to be made with respect to what to compute on the wearable and what to communicate to the fog, so as to balance the energy usage of the device.

\subsection{Mobility}
Driverless cars and drones are emerging application domains where fog computing plays a key role~\cite{chiang-2016}. Both these platforms have many on-board sensors and real-time analytics for autonomous mobility. Vehicular networks allow connected cars to communicate with each other (V2V) to cooperatively share information to make decisions on traffic and road conditions~\cite{hou-2016}. These can be extended to share compute capacity to perform these analytics. This allows a collection of parked or slow-moving vehicles to form a fog of resources among proximate ones. These may complement occasional road-side units that offer connectivity with the cloud. Here, the entities forming the fog can change dynamically, and requires distributed resource coordination.

Drones are finding use in asset monitoring, such as inspecting gas pipelines and power transmission towers at remote locations~\cite{loke-2015}. Here, a mobile base-station has a digital control tether with a collection of drones that follow a pre-set path to collect observational data. The drones have limited compute capacity to increase their endurance, and need to prioritize its use for autonomous navigation. So they typically serve as ``data mules'', collecting data from sensors along the route but doing limited on-board processing~\cite{mishra-2015}. However, when situations of interest arise, they can make use of their local capacity, resources available with near-by drones, or with fog resources in the base-station, while the trip is ongoing. This can help decide if a additional fly-past is necessary.


\section{Research Directions}

Fog computing is still exploratory in nature. While it has gained recent attention from researchers, many more topics need to be examined further.

\subsection{Fog Architectures} Many of the proposed fog architectural designs have not seen large scale deployments, and are just plausible proposals. While city-scale deployments with 100's of fog devices are coming online, the experiences from their operations will inform future design~\cite{yannuzzi-2017}. Network management is likely to be a key technical challenge as traffic management within the metropolitan area network (MAN) gains importance~\cite{vaquero-2014}. Unlike static fog resources, mobile or \emph{ad hoc} resources such as vehicular fog will pose challenges of resource discovery, access and coordination~\cite{hou-2016,he-2017}. Resource churn will need to be handled through intelligent scheduling~\cite{lopez-2015}. Resources will also require robust adaptation mechanisms based on the situational context~\cite{preden-2015a}. Open standards will be required to ensure interoperability. To this end, there are initial efforts on defining reference models for fog computing~\cite{openfog-2017}.

\subsection{Data Management} Tracking and managing content across edge, fog and cloud will be a key challenge. Part of this is the result of devices in IoT acting as data sources and compute platforms, which necessitates coordination across the cyber and physical worlds~\cite{anand-2017}. The generation of event streams that are transient and need to be processed in a timely manner poses additional challenges to the velocity dimension of big data~\cite{naas-2017}. Data discovery, replication, placement and persistence will need careful examination in the context of wide area networks and transient computing resources. Sensing will need to be complemented with ``sense-making'' so that data is interpreted correctly by integrating multiple sources~\cite{preden-2015a}.

\subsection{Programming Models and Platforms} Despite the growth of edge and fog resources, there is a lack of a common programming abstraction or runtime environments for defining and executing distributed applications on these resources
~\cite{stoica-2017}. There has been some preliminary work in defining a hierarchical pattern for composing applications that generate data from the edge and need to incrementally aggregate and process them at the fog and cloud layers~\cite{hong-2013}. They use spatial partitioning to assign edge devices to fogs. The ECHO platform offers a dataflow model to compose applications that are then scheduled on distributed runtime engines that are present on edge, fog and cloud resources~\cite{ravindra-2017}. It supports diverse execution engines such as NiFi, Storm and TensorFlow, but the user couples the tasks to a specific engine. A declarative application specification and big data platform is necessary to ease the composition of applications in such complex environments.

Related to this are application deployment and resource scheduling. VMs are used in the cloud to configure the required environment, but they may prove to be too resource intensive for fog. Some have examined the use of just a subset of the VM's footprint on the fog, and migrating this image across resources to track the mobility of user(s) who access its services~\cite{bittencourt-2015}. Resource scheduling on edge, fog and cloud have also been explored, though often validated just through simulations due to the lack of access to large scale fog setups~\cite{gupta-2017,zeng-2017}. Spatial awareness and energy awareness are distinctive features that have been included in such schedulers~\cite{brogi-2017,ghosh-2017}. Formal modeling of the fog has been undertaken as well~\cite{sarkar-2016}. Quality of Experience (QoE) as a user-centric alternative metric to Quality of Service (QoS) is also being examined~\cite{aazam-2016}. 

Such research will need to be revisited as the architectural and application models for fog computing become clearer, with mobility, availability and energy usage of resources offering unique challenges~\cite{shi-2012}.



\subsection{Security, Privacy, Trust} Unlike cloud computing where there is a degree of trust in the service provider, fog computing may contain resources from diverse and \emph{ad hoc} providers. Further, fog devices may not be physically secured like a data center, and may be accessible by third-parties~\cite{chiang-2016}. Containerization does not offer the same degree of sand-boxing between multiple tenant applications that virtualization does. Hence, data and applications in the fog operate within a mix of trusted and untrusted zones~\cite{lopez-2015}. This requires constant supervision of the device, fabric, applications and data by multiple stakeholders to ensure that security and privacy are not compromised. Techniques like anomaly detection, intrusion detection, moving target defense, etc. will need to be employed. Credential and identity management will be required. Provenance and auditing mechanisms will prove essential as well. These will need to be considered as first-class features when designing the fog deployment or the application.

\bibliographystyle{plain}  
\bibliography{chapter} 

\end{document}